\documentclass[prl,twocolumn,superscriptaddress,showpacs,preprintnumbers,amsmath,amssymb]{revtex4}
\usepackage{graphicx}
\usepackage{epsfig}
\usepackage{dcolumn}
\usepackage{bm}

\newcommand{\BR}{{\cal B}}

\newcommand{\piz}{\pi^0}

\newcommand{\jpsi}{J/\psi}

\newcommand{\EE}{e^+e^-}

\newcommand{\pp}{\pi^+\pi^-}
\newcommand{\kk}{K^+K^-}

\newcommand{\op}{\omega\phi}
\newcommand{\oo}{\omega\omega}
\newcommand{\etac}{\eta_c}
\newcommand{\ppp}{\pi^+\pi^-\pi^0}
\newcommand{\beq}{\begin{equation}}
\newcommand{\eeq}{\end{equation}}
\newcommand{\bitm}{\begin{itemize}}
\newcommand{\eitm}{\end{itemize}}

\begin{document}

\preprint{} \preprint{\vbox{ \hbox{   }
                        \hbox{Belle Preprint 2012-5}
                        \hbox{KEK   Preprint 2011-27}}}
\title{\quad\\[0.4cm]
Observation of new resonant structures in $\gamma \gamma \to \omega \phi$, $\phi
\phi$ and $\omega \omega$}

\affiliation{University of Bonn, Bonn}
\affiliation{Budker Institute of Nuclear Physics SB RAS and Novosibirsk State University, Novosibirsk 630090}
\affiliation{Faculty of Mathematics and Physics, Charles University, Prague}
\affiliation{Department of Physics, Fu Jen Catholic University, Taipei}
\affiliation{Justus-Liebig-Universit\"at Gie\ss{}en, Gie\ss{}en}
\affiliation{Gyeongsang National University, Chinju}
\affiliation{Hanyang University, Seoul}
\affiliation{University of Hawaii, Honolulu, Hawaii 96822}
\affiliation{High Energy Accelerator Research Organization (KEK), Tsukuba}
\affiliation{Hiroshima Institute of Technology, Hiroshima}
\affiliation{Indian Institute of Technology Guwahati, Guwahati}
\affiliation{Indian Institute of Technology Madras, Madras}
\affiliation{Institute of High Energy Physics, Chinese Academy of Sciences, Beijing}
\affiliation{Institute of High Energy Physics, Vienna}
\affiliation{Institute of High Energy Physics, Protvino}
\affiliation{INFN - Sezione di Torino, Torino}
\affiliation{Institute for Theoretical and Experimental Physics, Moscow}
\affiliation{J. Stefan Institute, Ljubljana}
\affiliation{Kanagawa University, Yokohama}
\affiliation{Institut f\"ur Experimentelle Kernphysik, Karlsruher Institut f\"ur Technologie, Karlsruhe}
\affiliation{Korea Institute of Science and Technology Information, Daejeon}
\affiliation{Korea University, Seoul}
\affiliation{Kyungpook National University, Taegu}
\affiliation{\'Ecole Polytechnique F\'ed\'erale de Lausanne (EPFL), Lausanne}
\affiliation{Faculty of Mathematics and Physics, University of Ljubljana, Ljubljana}
\affiliation{Luther College, Decorah, Iowa 52101}
\affiliation{University of Maribor, Maribor}
\affiliation{Max-Planck-Institut f\"ur Physik, M\"unchen}
\affiliation{University of Melbourne, School of Physics, Victoria 3010}
\affiliation{Graduate School of Science, Nagoya University, Nagoya}
\affiliation{Kobayashi-Maskawa Institute, Nagoya University, Nagoya}
\affiliation{Nara Women's University, Nara}
\affiliation{National Central University, Chung-li}
\affiliation{National United University, Miao Li}
\affiliation{Department of Physics, National Taiwan University, Taipei}
\affiliation{H. Niewodniczanski Institute of Nuclear Physics, Krakow}
\affiliation{Nippon Dental University, Niigata}
\affiliation{Niigata University, Niigata}
\affiliation{University of Nova Gorica, Nova Gorica}
\affiliation{Osaka City University, Osaka}
\affiliation{Pacific Northwest National Laboratory, Richland, Washington 99352}
\affiliation{Research Center for Nuclear Physics, Osaka University, Osaka}
\affiliation{RIKEN BNL Research Center, Upton, New York 11973}
\affiliation{University of Science and Technology of China, Hefei}
\affiliation{Seoul National University, Seoul}
\affiliation{Sungkyunkwan University, Suwon}
\affiliation{School of Physics, University of Sydney, NSW 2006}
\affiliation{Tata Institute of Fundamental Research, Mumbai}
\affiliation{Excellence Cluster Universe, Technische Universit\"at M\"unchen, Garching}
\affiliation{Toho University, Funabashi}
\affiliation{Tohoku Gakuin University, Tagajo}
\affiliation{Tohoku University, Sendai}
\affiliation{Department of Physics, University of Tokyo, Tokyo}
\affiliation{Tokyo Institute of Technology, Tokyo}
\affiliation{Tokyo Metropolitan University, Tokyo}
\affiliation{Tokyo University of Agriculture and Technology, Tokyo}
\affiliation{CNP, Virginia Polytechnic Institute and State University, Blacksburg, Virginia 24061}
\affiliation{Yamagata University, Yamagata}
\affiliation{Yonsei University, Seoul}
  \author{Z.~Q.~Liu}\affiliation{Institute of High Energy Physics, Chinese Academy of Sciences, Beijing} 
  \author{C.~P.~Shen}\affiliation{Graduate School of Science, Nagoya University, Nagoya} 
  \author{C.~Z.~Yuan}\affiliation{Institute of High Energy Physics, Chinese Academy of Sciences, Beijing} 
  \author{T.~Iijima}\affiliation{Kobayashi-Maskawa Institute, Nagoya University, Nagoya}\affiliation{Graduate School of Science, Nagoya University, Nagoya} 
  \author{I.~Adachi}\affiliation{High Energy Accelerator Research Organization (KEK), Tsukuba} 
  \author{H.~Aihara}\affiliation{Department of Physics, University of Tokyo, Tokyo} 
  \author{D.~M.~Asner}\affiliation{Pacific Northwest National Laboratory, Richland, Washington 99352} 
  \author{V.~Aulchenko}\affiliation{Budker Institute of Nuclear Physics SB RAS and Novosibirsk State University, Novosibirsk 630090} 
  \author{T.~Aushev}\affiliation{Institute for Theoretical and Experimental Physics, Moscow} 
  \author{A.~M.~Bakich}\affiliation{School of Physics, University of Sydney, NSW 2006} 
% ...
\collaboration{The Belle Collaboration}

\begin{abstract}

The processes $\gamma \gamma \to \omega \phi$, $\phi\phi$, and $\omega \omega$ are
measured using an 870~fb$^{-1}$ data sample collected with the
Belle detector at the KEKB asymmetric-energy $e^+e^-$ collider.
Production of vector meson pairs is clearly observed and their cross
sections are measured for masses that range from threshold to 4.0 GeV.
In addition to signals from well
established spin-zero and spin-two
charmonium states, there are resonant structures below charmonium threshold,
which have not been previously observed. We report a spin-parity analysis for the new structures
and determine the products of the $\etac$, $\chi_{c0}$, and $\chi_{c2}$
two-photon decay widths and branching fractions to
$\omega \phi$, $\phi\phi$, and $\omega \omega$.

\end{abstract}

\pacs{14.40.-n, 13.25.Gv, 13.25.Jx, 13.66.Bc}

\maketitle


A plethora of states, especially many new charmonium or
charmonium-like states (the so called ``$XYZ$ particles"),
that are not easily accommodated within
the quark model picture of hadrons have been observed~\cite{many}.
Recently a clear signal for a new state
$X(3915)\to \omega \jpsi$~\cite{x3915} and evidence for
another state $X(4350) \to \phi \jpsi$~\cite{x4350}
have been reported, thereby introducing new puzzles to
charmonium or charmonium-like spectroscopy.
Since these states couple
to a $\jpsi$ and a light mass vector, some authors have suggested
that they are good candidates for molecular or tetraquark
states~\cite{many}.

It is natural to extend the above theoretical picture to similar
states coupling to $\op$, since the only difference between such
states and the $X(3915)$~\cite{x3915} or $X(4350)$~\cite{x4350} is
the replacement of the $c\bar{c}$ pair with a pair of light quarks.
States coupling to $\oo$ or $\phi\phi$, although not as exotic as
those that decay into $\omega \phi$, which have two pairs of light quarks
in different generations, could also provide information on the
classification of the low-lying states coupled to pairs of light vector
mesons.

Experimental studies of  $\gamma \gamma \to V V$ ($V = \rho$,
$\omega$, $\phi$, $K^\ast$) began in 1980 with the measurement of
$\gamma \gamma \to \rho^0 \rho^0$~\cite{tasso}, and later $\gamma
\gamma \to \rho^+ \rho^-$~\cite{rhoprhom}. A number of theoretical
models, such as $q^2\bar{q}^2$ tetraquark states~\cite{fquark},
Regge exchange~\cite{regge}, and an $s$-channel $\rho^0 \rho^0$
resonance~\cite{schannel}, were proposed to explain
the large cross section observed in $\gamma \gamma \to \rho^0 \rho^0$
near the $\rho^0 \rho^0$ threshold that is absent in $\gamma
\gamma \to \rho^+ \rho^-$~\cite{vvreview}. The  $\gamma
\gamma \to \op$ and $\oo$ processes were studied by the ARGUS
Collaboration~\cite{argus:op,argus:oo} with very limited
statistics, while $\gamma \gamma \to \phi \phi$ has never been
measured below the charmonium mass region.

In this Letter, we report measurements of the cross sections for
$\gamma\gamma\to VV$, where $VV=\omega\phi, \phi\phi$ and
$\omega\omega$, as well as observations of new resonant structures
below charmonium threshold.
The results are based on an analysis of an 870~fb$^{-1}$ data sample taken at or
near the $\Upsilon(nS)$ ($n=1,...,5$) resonances with the Belle detector~\cite{Belle}
operating at the KEKB asymmetric-energy $\EE$ collider~\cite{KEKB}.
The Belle detector is described in detail elsewhere~\cite{Belle}.
We use the program {\sc treps}~\cite{treps} to generate signal
Monte Carlo (MC) events and determine experimental efficiencies
and luminosities.


We require four reconstructed charged tracks with zero net charge.
The selections of the charged kaon and pion tracks are the same as in Ref.~\cite{y1s}.
With this selection, the kaon (pion) identification efficiency is about 97\%
(98\%), while 0.4\% (1.0\%) of kaons (pions) are misidentified as
pions (kaons).
A similar likelihood ratio is formed for electron identification~\cite{EID}.
Photon conversion backgrounds are removed if any charged track in an
event is identified as electron or positron ($\mathcal{R}_e >
0.9$). For $\gamma \gamma \to \phi \phi$,
we require that only three of the charged
tracks be identified as kaons.

A good neutral cluster is reconstructed as a photon if its electromagnetic calorimeter (ECL)
shower does not match the extrapolation of any charged track and
its energy is greater than 50~MeV.
The $\pi^0$ candidates are reconstructed from pairs of photons with invariant mass within
15~MeV/$c^2$ of the $\piz$ nominal mass. Here the $\piz$ mass
resolution is about 6~MeV/$c^2$ from MC simulation. A
mass-constrained kinematic fit is applied to the selected $\pi^0$ candidate and
$\chi^2<10$ is required. For $\gamma \gamma \to \omega \omega$, the energies of
the photons from $\pi^0$ decays are further required to be greater
than 75~MeV in the endcap ECL region ($\cos\theta_{\gamma} < -0.65$)
to suppress  background with misreconstructed photons. When there
are more than two $\pi^0$ candidates in an event, the pair with
the smallest $\chi^2$ sum from the mass constraint is retained. To suppress
backgrounds with extra neutral clusters in the $\omega \phi$ and
$\omega \omega$ modes, events are removed if there are additional
photons with energy greater than 160 MeV.

We define the $\omega$ signal
region as 0.762~GeV/$c^2 <M(\ppp)< 0.802$~GeV/$c^2$, and the
$\omega$ mass sidebands region as 0.702~GeV/$c^2 <M(\ppp)<
0.742$~GeV/$c^2$ or 0.822~GeV/$c^2 <M(\ppp)< 0.862$~GeV/$c^2$,
which is twice as wide as the signal region. The $\phi$ signal
region is defined as 1.012~GeV/$c^2 <M(K^+K^-)< 1.027$~GeV/$c^2$,
and its sideband regions are defined as 0.99~GeV/$c^2<M(K^+K^-)<
1.005$~GeV/$c^2$ or 1.034~GeV/$c^2<M(K^+K^-)< 1.049$~GeV/$c^2$.
The $\phi$ sidebands are also twice as wide as the signal region.
The $VV$ pair sideband is defined as one $V$ in the signal region
while the other in the $V$ mass sideband.
For the two possible combinations of
$\phi \phi$ in the $2(K^+K^-)$ final state, the one with the smallest
$\delta_{min} = \sqrt{(M(\kk)_1-m_{\phi})^2 +
(M(\kk)_2-m_{\phi})^2}$ is chosen.
For the four possible combinations of $\omega \omega$,
only one combination from a true signal can survive after event selection.

\begin{figure}[htbp]
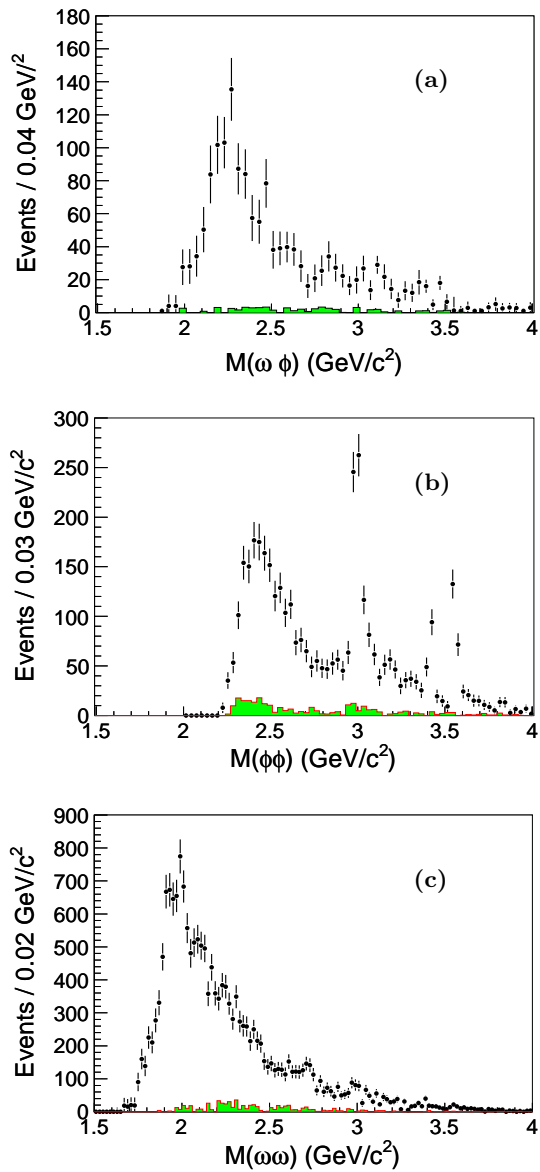

\psfig{file=fig1a.epsi,height=7.cm, angle=-90}
 \put(-50,-30){ \bf (a)}
 \\\bigskip
\psfig{file=fig1b.epsi,height=7.cm, angle=-90}
 \put(-50,-30){ \bf (b)}
 \\\bigskip
\psfig{file=fig1c.epsi,height=7.cm, angle=-90}
 \put(-50,-30){ \bf (c)}
\caption{ The (a) $\omega \phi$, (b) $\phi \phi$ and (c) $\omega \omega$ invariant mass distributions obtained by
fitting the $|\sum {\vec P}_t^{\ast}|$ distribution in each $VV$
mass bin. The shaded histograms are from the corresponding
normalized sidebands, which will be subtracted in calculating
the final cross sections.} \label{mass}
\end{figure}

The magnitude of the vector sum of the final particles' transverse
momenta in the $\EE$ center-of-mass (C.M.) frame, $|\sum {\vec
P}_t^{\ast}|$, which approximates the transverse momentum of the
two-photon-collision system, is used as a discriminating variable
to separate signal from background. The signal tends to
accumulate at small $|\sum {\vec P}_t^{\ast}|$ values while the
non-$\gamma \gamma$ background is distributed over a wider range.
We obtain the number of
$VV$ events in each $VV$ invariant mass bin by fitting the $|\sum
{\vec P}_t^{\ast}|$ distribution between zero and 0.9 GeV/$c$.
The signal shape is from MC simulation of the signal mode and the
background shape is parameterized as a second-order Chebyshev
polynomial. In order to control the background shape, we
restrict the coefficients of the background polynomials in nearby
invariant mass bins to vary smoothly.
The resulting $VV$ invariant mass
distributions are shown in Fig.~\ref{mass}.

There are some obvious structures in the low
$VV$ invariant mass region in Fig.~\ref{mass}. Two-dimensional (2D) angular distributions are
investigated to obtain the $J^P$ of the structures. In the process $\gamma
\gamma \to V V$, five angles are kinematically independent.
Among the possible variable sets, we choose $z$, $z^*$, $z^{**}$,
$\phi^*$, and $\phi^{**}$~\cite{angledef} and
use the transversity angle
($\phi_T$) and polar-angle product ($\Pi_\theta$) variables to analyze the
angular distributions. They are defined as \( \phi_T=|\phi^*+
\phi^{**}|/2\pi \), \( \Pi_\theta=[1-(z^*)^2][1-(z^{**})^2] \).

\begin{figure}[htbp]
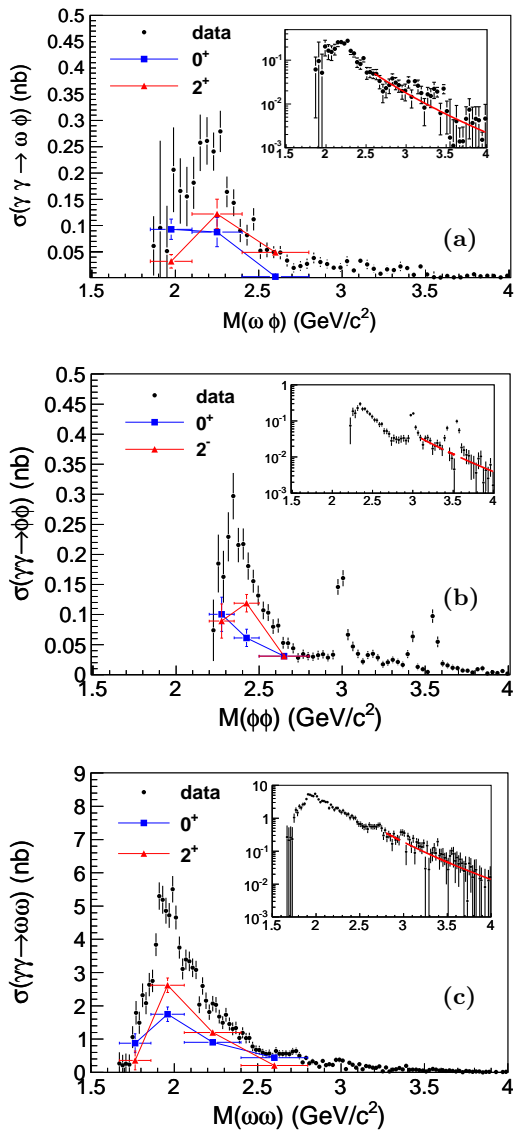

\includegraphics[height=6.7cm,angle=-90]{fig2a.epsi}
 \put(-30,-90){ \bf (a)}
\\\bigskip
\includegraphics[height=6.7cm,angle=-90]{fig2b.epsi}
 \put(-30,-90){ \bf (b)}
\\\bigskip
\includegraphics[height=6.7cm,angle=-90]{fig2c.epsi}
 \put(-30,-90){ \bf (c)}
\\\bigskip
\caption{The cross sections of $\gamma \gamma \to \omega \phi$
(a), $\phi \phi$ (b), and $\omega \omega$ (c)
are shown as points with error
bars. The cross sections for different $J^P$ values as a
function of $M(VV)$  are shown as the triangles and squares with error bars. For the processes $\gamma \gamma \to \omega
\phi$ and $\phi \phi$, the cross sections are measured in the C.M.
angular range $|\cos\theta^{\ast}|<0.8$, while for $\omega \omega$
the full $\hbox{cos}\theta^{\ast}$ range is covered.
 The error bars are
statistical only; there are overall systematic errors
of 15\%, 11\% and 13\% for $\omega \phi$,
$\phi \phi$ and $\omega \omega$, respectively.
The inset also shows the cross section on
a semi-logarithmic scale. In the high energy region, the solid curve
shows a fit to a $W^{-n}_{\gamma
\gamma}$  dependence  for the cross section after the significant charmonium contributions
($\eta_c$, $\chi_{c0}$ and $\chi_{c2}$) were excluded. }
\label{cross-section}
\end{figure}

We obtain the number of signal events by fitting the $|\sum {\vec P}_t^{\ast}|$
distribution in each $\phi_T$ and $\Pi_\theta$ bin in the 2D
space, which is divided into $4\times 4$, $5\times 5$, and
$10\times 10$ bins for $\omega \phi$, $\phi \phi$, and $\omega
\omega$, respectively, for $M(VV)<2.8$~GeV/$c^2$,
in some wider $VV$ mass bins as shown in Fig.~\ref{cross-section}.
The obtained 2D angular distribution data are fitted with the
signal shapes from MC-simulated samples with different $J^P$
assumptions ($0^+$, $0^-$, $2^+$, $2^-$). We find:
(1) for $\omega \phi$: $0^+$ ($S$-wave) or $2^+$ ($S$-wave)
can describe data with $\chi^2/ndf=1.1$ or 1.2, while
  a mixture of $0^+$ ($S$-wave)
and $2^+$ ($S$-wave) describes data with $\chi^2/ndf=0.9$ ($ndf$
is the number of degrees of freedom); (2) for $\phi \phi$: a mixture
of $0^+$ ($S$-wave) and  $2^-$ ($P$-wave) describes data with
$\chi^2/ndf=1.3$; and (3) for $\omega \omega$: a mixture of
$0^+$ ($S$-wave) and $2^+$ ($S$-wave) describes data with
$\chi^2/ndf=1.3$.
The contributions from other $J^P$ are found to be small
and thus neglected.

The cross section $\sigma_{\gamma \gamma \to VV}(W_{\gamma
\gamma})$ is calculated from
\begin{equation}\label{csfor}
\sigma_{\gamma \gamma \to VV}(W_{\gamma \gamma}) = \frac{\Delta
n}{\frac{dL_{\gamma \gamma}}{dW_{\gamma \gamma}}
\epsilon(W_{\gamma \gamma}) \Delta W_{\gamma \gamma}},
\end{equation}
where  $\frac{dL_{\gamma \gamma}}{dW_{\gamma \gamma}}$ is the
differential luminosity of the two-photon collision, and
$\epsilon$ is the efficiency. Here $\Delta W_{\gamma \gamma}$ is the
bin width and $\Delta n$ is the number of events in the $\Delta
W_{\gamma \gamma}$ bin.

The $\gamma \gamma \to VV$ cross sections are shown in
Fig.~\ref{cross-section}. For the processes $\gamma \gamma
\to \omega \phi$ and $\phi \phi$, the cross sections are measured
in the C.M. angular range $|\hbox{cos}\theta^{\ast}|<0.8$ since
there are no detected events beyond this limit, while for $\omega \omega$
the full $\hbox{cos}\theta^{\ast}$ range is covered.
The cross sections for
different $J^P$ values as a function of $M(VV)$ are also shown in
Fig.~\ref{cross-section}.
We observe structures
at  $M(\omega \phi)\sim 2.2$~GeV/$c^2$, $M(\phi \phi)\sim
2.35$~GeV/$c^2$, and $M(\omega \omega)\sim 1.91$~GeV/$c^2$ with peak cross
sections of ($0.27\pm 0.05$)~nb, $(0.30\pm 0.04)$~nb, and $(5.30\pm 0.42)$~nb, respectively. While there are substantial
spin-zero components in all three modes, there are also
significant spin-two components, at least in the $\phi\phi$ and $\omega\omega$ modes.
The phase space enhancement effect should be much closer to the $VV$ mass thresholds and
it is impossible to produce the observed mass-dependent cross sections.

The inset also shows the distribution of the cross
section on a semi-logarithmic scale, where, in the high
mass region, we fit the $W^{-n}_{\gamma \gamma}$ dependence of
the cross section. The solid curves are the fitted
results; the fit gives $n=7.2\pm0.6$, $8.4\pm1.1$, and $9.1\pm0.6$ for the $\omega \phi$,
$\omega \omega$, and $\phi \phi$ modes, respectively. These
results are consistent with the predictions from pQCD and handbag models~\cite{handbag},
and similar to previous measurements in other
modes~\cite{nn}.

There are several sources of systematic error for the
cross section measurements. The particle identification
uncertainties are 1.5\% for each kaon, 1.2\% for each pion.
A momentum-weighted systematic error in tracking efficiency is
taken for each track, which is about 0.6\%. The efficiency uncertainties associated with
the $\omega$ and $\phi$ mass requirements are almost independent
of the $VV$ mass, and are estimated to be 1.9\% and 1.6\%, respectively. The
statistical error in the MC samples is about 0.5\%. The accuracy
of the two-photon luminosity function calculated with the {\sc
treps} generator is estimated to be about 5\% including the error
from neglecting radiative corrections (2\%), the uncertainty from
the form factor effect (2\%), and the uncertainty in the total
integrated luminosity (1.4\%)~\cite{treps}.
The uncertainty of
the trigger simulation is smaller than 5\%~\cite{epjc}. The
preselection efficiency for the final states has little dependence
on the $VV$ invariant mass, with an uncertainty that is smaller
than 1\% for $\omega \phi$, 4\% for $\phi \phi$ and 2.5\% for
$\omega \omega$. From Ref.~\cite{PDG}, the uncertainty in the
world average values for $\BR(\phi\to \kk)$ is 1.1\% and that for
$\BR(\omega\to \pp \pi^0)$ is 0.8\%.
The uncertainty in the fitted yield for the signal is estimated by
varying the order of the background polynomial and fit range,
which is 10\% for $\omega \phi$, 2.5\% for $\phi \phi$, and 4.0\%
for $\omega \omega$. The uncertainty on the $|\sum {\vec
P}_t^{\ast}|$ resolution is smaller than 2.2\%, which is estimated
by changing the MC signal resolution by $\pm 10\%$. The
uncertainty on the weighted efficiency curve is estimated by
changing the fitted ratio of the $J^P$ components by $\pm
1\sigma$, which is 1.0\% for $\omega \phi$, 3.1\% for $\phi \phi$,
and 1.0\% for $\omega \omega$. Assuming that
all of these systematic error sources are independent, we add them
in quadrature to obtain the total systematic errors, which are
15\%, 11\% and 13\% for $\omega \phi$, $\phi \phi$ and $\omega \omega$, respectively.

For $VV$ invariant masses above 2.8~GeV/$c^2$, we measure the
production rate of charmonium states.
In measuring the production rates, $|\sum
{\vec P}_t^{\ast}|$ is required to be less than $0.1~\hbox{GeV}/c$
in order to reduce backgrounds from non-two-photon-processes and
two-photon-processes with extra particles other than $\phi$ or
$\omega$ in the final state.

\begin{figure}[htbp]
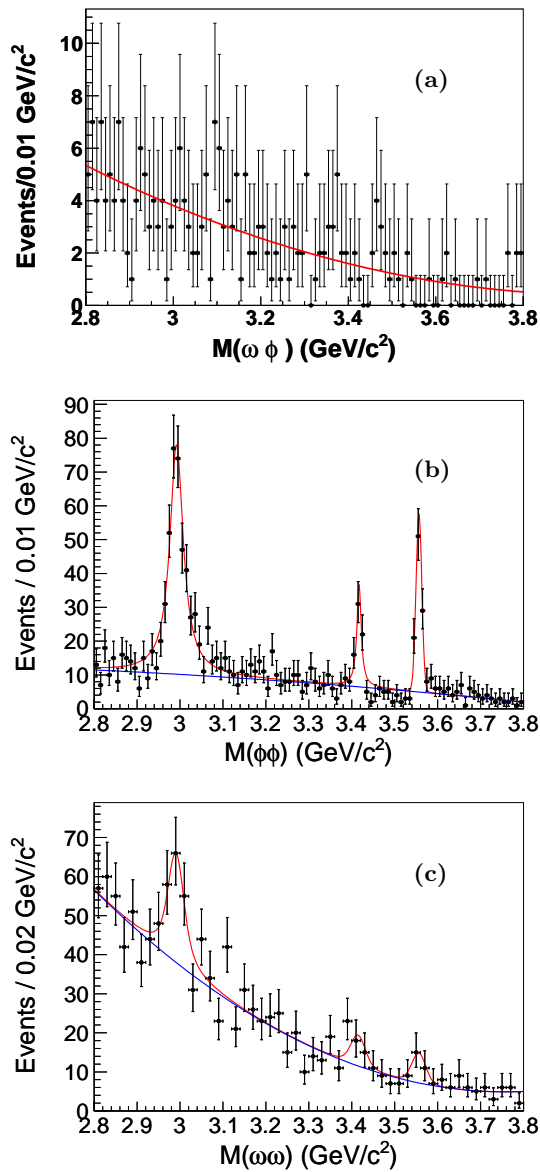

\includegraphics[height=7cm,angle=-90]{fig3a.epsi}
 \put(-50,-30){ \bf (a)}
\\\bigskip
\includegraphics[height=7cm,angle=-90]{fig3b.epsi}
 \put(-50,-30){ \bf (b)}
\\\bigskip
\includegraphics[height=7cm,angle=-90]{fig3c.epsi}
 \put(-50,-30){ \bf (c)}
\\\bigskip
\caption{ The invariant mass distributions of (a) $\op$, (b)
$\phi\phi$, and (c) $\oo$ combinations in the charmonium mass region with the requirement of  $|\sum {\vec
P}_t^{\ast}|<0.1~\hbox{GeV}/c$. The points
with error bars are data, and the solid curves are the best fits.}
\label{charmonium}
\end{figure}

Figure~\ref{charmonium} shows the $VV$ invariant mass
distributions and best fits. Clear $\etac$, $\chi_{c0}$ and $\chi_{c2}\to \phi
\phi$, and $\eta_c\to \oo$ signals are evident. The $VV$ mass distributions
are fitted with three incoherent Breit-Wigner functions convoluted
with a corresponding double Gaussian resolution function as the
$\etac$, $\chi_{c0}$ and $\chi_{c2}$ signal shapes, and a
second-order Chebyshev polynomial as the background shape.


The numbers of signal events and
product of the two-photon decay width and branching fraction
$\Gamma_{\gamma\gamma}\BR(X \to VV)$ (or the upper limits in case
the signal is insignificant) for $\etac$, $\chi_{c0}$ and
$\chi_{c2}$ are listed in Table~\ref{bgamfinal}.
In these calculations,
we assume there is no interference between the charmonium and the
continuum amplitudes~\cite{coherent}. A systematic error estimate similar to that for the cross sections
considers additionally the uncertainties on the resonance parameters
and results in the total systematic errors of 13\%, 11\%, and 11\% for
$\Gamma_{\gamma \gamma}(R) \BR(R \to \op)$; 7.9\%, 8.0\%, and
7.2\% for $\Gamma_{\gamma \gamma}(R) \BR(R \to \phi \phi)$; and
11\%, 10\%, and 9.1\% for $\Gamma_{\gamma \gamma}(R) \BR(R \to
\omega \omega)$, for $R=\etac$, $\chi_{c0}$ and $\chi_{c2}$,
respectively. For the upper limit determinations, the efficiencies
have been lowered by a factor of $1-\sigma_{\rm sys}$ in order to
obtain conservative values. The measurements of
$\Gamma_{\gamma\gamma}\BR(X \to\phi \phi)$ for $\etac$,
$\chi_{c0}$ and $\chi_{c2}$ are consistent with previously
published results~\cite{epjc} with improved precision. The values
of  $\Gamma_{\gamma\gamma}\BR(X \to\phi \phi)$ for $\etac$,
$\chi_{c0}$ and $\chi_{c2}$  obtained in this work supersede those
in Ref.~\cite{epjc}. All the other results are first
measurements.

\begin{table*}[htbp]
\caption{ Results for  $\Gamma_{\gamma\gamma}\BR(X \to VV)$ (eV)
and the numbers of events (in brackets) for
$\etac$, $\chi_{c0}$ and $\chi_{c2}$, where the values of
$\BR(\omega \to \pi^+ \pi^- \pi^0)=(89.2\pm 0.7)\%$ and $\BR(\phi
\to K^+ K^-) = (48.9\pm 0.5)\%$ are used~\cite{PDG}. The first and
second errors for the central values are statistical and
systematic, respectively. The upper limits are obtained at the 90\% confidence level
with systematic errors included.} \label{bgamfinal}
\begin{center}
\begin{tabular}{c  c  c  c}
 \hline
 Mode          & $\omega \phi$  & $\phi \phi$ & $\omega \omega$  \\\hline
 $\etac$ &   $<0.49$ [$<7.9$]  & $7.75\pm0.66\pm0.62$ [$386\pm31$] & $8.67\pm2.86\pm 0.96$ [$85\pm29$]  \\
 $\chi_{c0}$   & $<0.34$ [$<4.3$]  & ~$1.72\pm0.33\pm 0.14$~ [$56\pm11$]  & $<3.9$ [$<35$]  \\
 $\chi_{c2}$ & $<0.04$ [$<2.4$] & $0.62\pm0.07\pm0.05$ [$89\pm11$]  & $<0.64$ [$<28$]   \\
 \hline
\end{tabular}
\end{center}
\end{table*}


In summary, we present a search for exotic states in two-photon processes
$\gamma \gamma \to \op$, $\phi \phi$ and $\oo$. The production of $\op$,  $\phi \phi$,
and $\oo$ is observed, and cross sections are measured up to 4 GeV/$c^2$.
The cross sections for $\gamma \gamma \to \omega \phi$ are
much lower than the prediction of the $q^2 \bar{q}^2$ tetraquark model~\cite{vvreview}
of 1~nb, while the resonant structure in the $\gamma \gamma \to \phi \phi$ mode is
nearly at the predicted position. However, the $\phi \phi$ cross section
is an order of magnitude lower than the expectation in the tetraquark model.
On the other hand, the t-channel factorization model~\cite{alexander} predicted that
the $\phi \phi$ cross sections vary between 0.001~nb and 0.05~nb in the mass region
of 2.0 GeV/$c^2$ to 5.0 GeV/$c^2$, which are much lower than the experimental data.
For $\gamma \gamma \to \omega \omega$, the t-channel factorization model~\cite{alexander}
predicted a broad structure between 1.8 GeV/$c^2$ and 3.0 GeV/$c^2$ with a peak cross section
of 10-30~nb near 2.2 GeV/$c^2$, while the one-pion-exchange model~\cite{AKS} predicted an
enhancement near threshold around 1.6 GeV/$c^2$ with a peak cross section of 13~nb using
a preferred value of the slope parameter.
Both the peak position and the peak height predicted in~\cite{alexander} and~\cite{AKS}
disagree with our measurements.


We thank the KEKB group for excellent operation of the
accelerator; the KEK cryogenics group for efficient solenoid
operations; and the KEK computer group, the NII, and
PNNL/EMSL for valuable computing and SINET4 network support.
We acknowledge support from MEXT, JSPS and Nagoya's TLPRC (Japan);
ARC and DIISR (Australia); NSFC (China); MSMT (Czechia);
DST (India); INFN (Italy); MEST, NRF, GSDC of KISTI, and WCU (Korea);
MNiSW (Poland); MES and RFAAE (Russia); ARRS (Slovenia);
SNSF (Switzerland); NSC and MOE (Taiwan); and DOE and NSF (USA).



\begin{thebibliography}{**}
\bibitem{coherent} Since the $\eta_c  \to \phi \phi$ signal
is significant and the $\eta_c$ width is relatively large, we also fit
with a coherent BW function for the $\eta_c$. There are two solutions
with equally good fit quality (the goodness of the fit is
$\chi^2/ndf=0.6$). The products of the two-photon decay width and
branching fraction $\Gamma_{\gamma\gamma}\BR(\eta_c \to \phi
\phi)$ are $3.13\pm 0.55$~eV for the constructive solution, and
$15.7\pm 1.9$~eV for the destructive solution, where the errors
are statistical only.
\end{thebibliography}
\end{document}